\documentstyle[preprint,aps,epsf,floats]{revtex}
\begin{document}
\tighten

\def\bfl{{\bbox \ell}}
\def\bull{\vrule height .9ex width .8ex depth -.1ex}
\def\MeV{{\rm MeV}}
\def\GeV{{\rm GeV}}
\def\Tr{{\rm Tr\,}}
\def\nrcpt{NR\raise.4ex\hbox{$\chi$}PT\ }
\def\ket#1{\vert#1\rangle}
\def\bra#1{\langle#1\vert}
\def\ltap{\ \raise.3ex\hbox{$<$\kern-.75em\lower1ex\hbox{$\sim$}}\ }
\def\gtap{\ \raise.3ex\hbox{$>$\kern-.75em\lower1ex\hbox{$\sim$}}\ }
\def\abs#1{\left| #1\right|}
\def\CA{{\cal A}}
\def\CC{{\cal C}}
\def\CD{{\cal D}}
\def\CE{{\cal E}}
\def\CL{{\cal L}}
\def\CO{{\cal O}}
\def\CZ{{\cal Z}}
\def\bvert{\Bigl\vert\Bigr.}
\def\pds{{\it PDS}\ }
\def\ms{MS}
\def\ddq{{{\rm d}^dq \over (2\pi)^d}\,}
\def\ddqm{{{\rm d}^{d-1}{\bf q} \over (2\pi)^{d-1}}\,}
\def\bfq{{\bf q}}
\def\bfk{{\bf k}}
\def\bfp{{\bf p}}
\def\bfpp{{\bf p '}}
\def\bfr{{\bf r}}
\def\dtr{{\rm d}^3\bfr\,}
\def\bfx{{\bf x}}
\def\dtx{{\rm d}^3\bfx\,}
\def\dfx{{\rm d}^4 x\,}
\def\bfy{{\bf y}}
\def\dty{{\rm d}^3\bfy\,}
\def\dfy{{\rm d}^4 y\,}
\def\dfq{{{\rm d}^4 q\over (2\pi)^4}\,}
\def\dfk{{{\rm d}^4 k\over (2\pi)^4}\,}
\def\dfl{{{\rm d}^4 \ell\over (2\pi)^4}\,}
\def\dtq{{{\rm d}^3 {\bf q}\over (2\pi)^3}\,}
\def\dtk{{{\rm d}^3 {\bf k}\over (2\pi)^3}\,}
\def\dtl{{{\rm d}^3 {\bfl}\over (2\pi)^3}\,}
\def\dt{{\rm d}t\,}
\def\frac#1#2{{\textstyle{#1\over#2}}}
\def\darr#1{\raise1.5ex\hbox{$\leftrightarrow$}\mkern-16.5mu #1}
\def\){\right)}
\def\({\left( }
\def\]{\right] }
\def\[{\left[ }
\def\si{{}^1\kern-.14em S_0}
\def\siii{{}^3\kern-.14em S_1}
\def\diii{{}^3\kern-.14em D_1}
\def\p0iii{{}^3\kern-.14em P_0}
\def\fm{{\rm\ fm}}
\def\MeV{{\rm\ MeV}}
\def\CA{{\cal A}}
\def\Czzm{ {\cal A}_{-1[00]} }
\def\Cttm{{\cal A}_{-1[22]} }
\def\Ctzm{{\cal A}_{-1[20]} }
\def\Cztm{ {\cal A}_{-1[02]} }
\def\Czzz{{\cal A}_{0[00]} }
\def\Cttz{ {\cal A}_{0[22]} }
\def\Ctzz{{\cal A}_{0[20]} }
\def\Cztz{{\cal A}_{0[02]} }


\def\spzz{ {Y_{sp0}^{(0)} }}
\def\spzzB{ {Z_{sp0}^{(0)} }}
\def\spzo{ {Y_{sp0}^{(1)} }}
\def\spzt{ {Y_{sp0}^{(2)} }}
\def\ppspz{ {y^{sp0}_0 }}
\def\ppspt{ {y^{sp0}_2 }}
\def\ppspB{ {z^{sp0}_2 }}

\def\Ames{ A }  

\newcommand{\eqn}[1]{\label{eq:#1}}
\newcommand{\refeq}[1]{(\ref{eq:#1})}
\newcommand{\eq}{eq.~\refeq}
\newcommand{\eqs}[2]{eqs.~(\ref{eq:#1}-\ref{eq:#2})}
\newcommand{\eqsii}[2]{eqs.~(\ref{eq:#1}, \ref{eq:#2})}
\newcommand{\Eq}{Eq.~\refeq}
\newcommand{\Eqs}{Eqs.~\refeq}

\def\Journal#1#2#3#4{{#1} {\bf #2}, #3 (#4)}

\def\NCA{\em Nuovo Cimento}
\def\NIM{\em Nucl. Instrum. Methods}
\def\NIMA{{\em Nucl. Instrum. Methods} A}
\def\NPB{{\em Nucl. Phys.} B}
\def\NPA{{\em Nucl. Phys.} A}
\def\PLB{{\em Phys. Lett.}  B}
\def\PRL{\em Phys. Rev. Lett.}
\def\PRD{{\em Phys. Rev.} D}
\def\PRC{{\em Phys. Rev.} C}
\def\ZPC{{\em Z. Phys.} C}
\def\PPN{{\em Prog. Part. Nucl. Phys.} }
\def\ARNPS{{\em Ann. Rev. Nucl. Part. Sci.} }
\def\CNPP{{\em Comments Nucl. Part. Phys.} }
\def\NPPS{{\em Nucl. Phys. Proc. Suppl.} }
\def\PRep{{\em Phys. Rep.} }
\def\AFA{{\em Ann. Fis.} A}
\def\JPC{{\em J. Phys.} C3}
\def\PTPS{{\em Prog. Theor. Phys. Suppl.} }
\def\PTP{{\em Prog. Theor. Phys.} }

\preprint{\vbox{
\hbox{ NT@UW-99-02}
}}
\bigskip
\bigskip

\title{Pionic Matrix Elements in Neutrinoless Double-$\beta$ Decay}
\author{Martin J. Savage}
\address{ Department of Physics, University of Washington,  
Seattle, WA 98915 }
\address{and\ Jefferson Lab., 12000 Jefferson Avenue, Newport News, 
Virginia 23606
\\ {\tt savage@phys.washington.edu} }  
\maketitle


\begin{abstract}
Physics beyond the standard model of electroweak interactions that violates
lepton number will, in general,
induce operators involving the quark and lepton fields that
violate lepton number.
I point out that some matrix elements of particular operators are known.
In the limit of exact SU(3) flavor symmetry the operator
$\overline{u}\gamma_\mu (1-\gamma_5)d \ \overline{u}\gamma_\mu (1-\gamma_5)d$
is part of the ${\bf 27}$ representation responsible for
$\Delta~I~=~{3\over 2}$ strangeness changing decays in the standard model.
At leading order in the chiral expansion there is only one coupling constant
that describes the weak interactions of the pseudo-Goldstone bosons
induced by the  ${\bf 27}$ representation.
The observed branching fraction for 
$K^+\rightarrow\pi^+\pi^0$ determines this coupling constant.
Contributions of the form $m_s\log m_s$ are used to estimate
SU(3) breaking corrections to the relation between 
$K^+\rightarrow\pi^+\pi^0$ and
the $\pi^-\rightarrow \pi^+ e^-e^-$ interaction.
\end{abstract}

\vskip 2in

\leftline{November 1998}
%
%
%
%
\vfill\eject

Observation of neutrinoless double-$\beta$ decay  ($\beta\beta_{0\nu}$)
between two nuclei,  $(A,Z)\rightarrow (A,Z+2) + e^- + e^-$,
would provide unambiguous evidence of physics beyond the standard model,
due to the explicit violation lepton number.
Ongoing experimental efforts to find evidence for such transitions have so far
been unsuccessful.
While cleanly observing the lepton number conserving two-neutrino
($\beta\beta_{2\nu}$) mode
$(A,Z)\rightarrow (A,Z+2) + e^- + e^- + \overline{\nu}_e + \overline{\nu}_e$
that arises at second order in the weak interaction,
only lower limits on the lifetime for  the neutrinoless process
$(A,Z)\rightarrow (A,Z+2) + e^- + e^-$
have been presented(for a recent review see \cite{Morales}).

Forming theoretical predictions for the rate of both the 
$\beta\beta_{2\nu}$ and $\beta\beta_{0\nu}$
decay processes presents challenges in many areas.
The underlying physics responsible for $\beta\beta_{2\nu}$ decays
is (in the standard model)
simply two charged current weak interactions, dominated by operators involving
only the {\it up} and {\it down}  quarks.
The matrix element of each of these weak interactions in the nucleon and pion
are well
known from ordinary $\beta$-decay and therefore 
the challenge arises in forming the nuclear matrix elements of the time-ordered
product of these two weak interaction.
Predictions of nuclear model calculations for these $\beta\beta_{2\nu}$ decay
rates  tend to vary by factors
$\sim 5$\cite{Vergados,Wickb,Kotani,Caurier,Retamosa,Vogel,Faessler,Klapdor,Suhonen,Wu,Hirsch,Abad}.

In contrast, physics beyond the standard model that might give rise to
lepton number non-conservation is unknown.
Its existence, origin and structure can only be speculated.
Extensive theoretical study in this area has focused on lepton number violation in the
neutrino mass matrix,
and in particular lepton number violating interactions
inserted into the propagator of very light neutrinos
(an excellent review  by Haxton and Stephenson
can be found in \cite{Wick}).
While one expects insertions of the lowest dimension operators to make the
dominant contribution, it is possible that significant lepton number violation occurs
via the exchange of heavy particles.
At low energies,
such exchanges will induce local operators involving the quark and lepton
fields of the standard model.  
Historically, specific models such as
left-right-symmetric models\cite{Mohap}, 
supersymmetry with R-parity violation\cite{susy},
or other lepton number violating extensions to the standard model\cite{maj},
have been examined.
The lepton number violating operators
resulting from such scenarios have been determined
perturbatively in the short-distance coupling constants.
It is perhaps not surprising that difficulties
arise in evaluating the hadronic matrix elements
of the resulting operators
between nucleon and pion fields.
Factorization has been extensively employed to estimate 
the hadronic matrix elements of multi-quark operators.
However,  it is highly desirable to make more reliable estimates.
In addition, the difficulties encountered in
evaluating the nuclear matrix elements of $\beta\beta_{2\nu}$ decays
must still be dealt with in evaluating the rates for $\beta\beta_{0\nu}$
processes.

In this work I point out the matrix element of a certain operator structure can
be related to the amplitudes for $\Delta I={3\over 2}$ processes in the SU(3)
limit.
Restricting ourselves to discussions of operators involving four quark fields
and two electron fields, the most general lepton number violating
operator involving electrons and first
generation quarks is
\begin{eqnarray}
  {\cal O}^{\Delta L=2} & = & 
\overline{u} \Gamma_q d \ \ 
\overline{u} \Gamma_q d \ \ 
\overline{e}\Gamma_l e^c
\ \ \ ,
\end{eqnarray}
where $e^c$ is the charge conjugate electron field, $\Gamma_l$ is the lorentz
structure responsible for the lepton interaction and $\Gamma_q$ is the lorentz structure
responsible for the quark interaction.
Of the many forms of $\Gamma_q$, matrix elements of
operators with
$\Gamma_q = \gamma_\mu (1-\gamma_5)$ or $\gamma_\mu (1+\gamma_5)$
can be related by SU(3) symmetry
to matrix elements of
\begin{eqnarray}
  {\cal O}^{\Delta s=1} & = & 
\overline{u}\gamma_\mu (1-\gamma_5) s\
\overline{d}\gamma^\mu (1-\gamma_5) u
\ \ \ ,
\end{eqnarray}
that arises in the standard model.

In the standard model the exchange of a single $W$-gauge boson 
induces an effective lagrange density
\begin{eqnarray}
  {\cal L}^{\Delta s=1} & = &
  - {G_F\over\sqrt{2}} \ V_{us}V_{ud}^*\ 
\overline{u}\gamma_\mu (1-\gamma_5) s\
\overline{d}\gamma^\mu (1-\gamma_5) u
\ +\ ...
\nonumber\\
& = &
 - {G_F\over\sqrt{2}} \ V_{us}V_{ud}^*\
 \left[ C^{\bf (8)}(M_W) {\cal O}^{\bf (8)}(M_W)
   \ +\ 
 C^{\bf (27)}(M_W) {\cal O}^{\bf (27)}(M_W)\right]
\ \ \ ,
\end{eqnarray}
where the ellipsis denote higher dimension operators suppressed by
powers of the $W$-boson mass.
$G_F$ is Fermi's coupling constant and $V_{ij}$ is the element of the
Cabibbo-Kobayashi-Maskawa (CKM) weak mixing matrix between 
quarks of flavor $i$ and $j$.
Under SU(3) flavor symmetry  $ {\cal L}^{\Delta s=1}$ 
can be decomposed into operators ,
${\cal O}^{\bf (8)}$ and ${\cal O}^{\bf (27)}$,
that transform as ${\bf 8}\oplus {\bf 27}$ respectively.
The coefficients of the operators, $C^{\bf (8)}(\mu)$ and 
$C^{\bf (27)}(\mu)$,  are renormalized at the scale $\mu$.
It is well known that these 
irreducible representations are renormalized differently
by the strong interaction, with the ${\bf 8}$
being enhanced in the infrared while the ${\bf 27}$ is suppressed.
The ${\bf 27}$ is not renormalized by peguin-type contributions,
and in scaling between  
renormalization scales $\mu$ and $\mu^\prime$ one finds  in
the leading log approximation
\begin{eqnarray}
 C^{\bf (27)}(\mu^\prime) & = &  C^{\bf (27)}(\mu)
 \left[{\alpha_s(\mu^\prime)\over \alpha_s(\mu)}\right]^{-{2\over b}}
\ \ \ ,
\end{eqnarray}
where $b=11-{2\over 3} N_f$ with $N_f$ being the number of active quarks
between the scales $\mu$ and $\mu^\prime$ and
$\alpha_s (\mu)$ is the strong interaction coupling constant at the scale
$\mu$.

In flavor space, suppressing strong interaction indices, 
the operator ${\cal O}^{\bf (27)}$ is
\begin{eqnarray}
  {\cal O}^{\bf (27)} & = &  T^{ab}_{cd} \ 
\overline{q}^c\gamma_\mu (1-\gamma_5) q_a\
\overline{q}^d\gamma^\mu (1-\gamma_5) q_b
\nonumber\\
T^{13}_{12} = T^{13}_{21} = T^{31}_{12} = T^{31}_{21} & = & {1\over 5}
\ \ ,\ \
T^{23}_{22} = T^{32}_{22} = T^{33}_{32} = T^{33}_{23} = -{1\over 10}
\ \ \ ,
\label{tsop}
\end{eqnarray}
and when renormalized at the weak scale ($\mu=M_W$) 
its Wilson coefficient is $C^{\bf (27)}(M_W)=1$.
Under chiral $SU(3)_L\otimes SU(3)_R$ chiral transformations the tensor
$T$ transforms as
$T^{ab}_{cd}\rightarrow L^a_\alpha L^b_\beta T^{\alpha\beta}_{\sigma\rho}
L^{\dagger\sigma}_c L^{\dagger\rho}_d$, and 
matrix elements of ${\cal O}^{\bf (27)}$ between meson states are reproduced by
an effective lagrange density (at leading order) 
\begin{eqnarray}
  \langle \pi' s | C^{\bf (27)}(M_W) {\cal O}^{\bf (27)}(M_W) |\pi' s\rangle 
  & = &
  \langle \pi' s |\
  g^{\bf (27)}\ f_\pi^4\ T^{ab}_{cd}
  \left(\Sigma\partial_\mu\Sigma^\dagger\right)^c_a
  \left(\Sigma\partial^\mu\Sigma^\dagger\right)^d_b
  \ +\ ...\ |\pi' s\rangle
\ \ \ ,
\label{sigmat}
\end{eqnarray}
where $f_\pi = 131\ {\rm MeV}$ is the pion decay constant.
The ellipsis denote operators involving more derivatives acting on the meson
field and insertions of the light quark mass matrix.
$\Sigma$ is the exponential of the pseudo-Goldstone boson field,
\begin{eqnarray}
  \Sigma & = & \exp\left({2 i\over f_\pi} M\right)
  \ \ ,\ \
  M = \left( \matrix{\pi^0/\sqrt{2} + \eta/\sqrt{6} & \pi^+ & K^+ \cr
      \pi^- & -\pi^0/\sqrt{2} + \eta/\sqrt{6} & K^0 \cr
      K^- & \overline{K}^0 & -2/\sqrt{6}\eta }\right)
\ \ \ .
\end{eqnarray}
The experimentally observed rate for $K^+\rightarrow \pi^+\pi^0$ fixes
$g^{\bf (27)} = 0.12$ (see \cite{Valen}) in the limit of SU(3)
symmetry\footnote{Our definitions of $f_\pi$ and $T^{ab}_{cd}$
  lead to $g^{\bf (27)}$ being  a factor of
  ${3\over 4}$ smaller than that used in \cite{Valen}.}
while naive dimensional analysis would suggest $g^{\bf (27)} \sim 1$.
This suppression is part of the $\Delta I={1\over 2}$ rule one finds in
$\Delta s=1$ decays.

Below the scale of physics responsible for lepton number violation $\Lambda$,
one may find an $\Delta L=2$
effective lagrange density involving the quark and lepton fields
of the form
\begin{eqnarray}
 {\cal L}_{\rm \beta\beta}
  & = &  
   C_{\beta\beta} (\Lambda) \ 
   \overline{u}\gamma_\mu (1-\gamma_5) d\
  \overline{u}\gamma^\mu (1-\gamma_5) d \  \overline{e}\Gamma_l e^c\ (\Lambda)
  \nonumber\\
  & = & C_{\beta\beta} (\Lambda) \  T^{ab}_{cd} \ 
  \overline{q}^c\gamma_\mu (1-\gamma_5) q_a\
  \overline{q}^d\gamma^\mu (1-\gamma_5) q_b  \  \overline{e}\Gamma_l e^c\
  (\Lambda)
  \nonumber\\
T^{22}_{11} & = & 1
\ \ \ .
\label{doubleop}
\end{eqnarray}
This operator transforms is a component of the  ${\bf 27}$ of SU(3),
and SU(3) symmetry relates matrix elements of
this operator to those of the $ {\cal O}^{\bf (27)}$ in eq.~(\ref{tsop}).
The effective lagrange density that reproduces 
matrix elements between meson states is found from eq.~(\ref{sigmat})
to be
\begin{eqnarray}
   {\cal L}_{\rm \beta\beta}
& \rightarrow &
 C_{\beta\beta} (M_W)\ 
g^{\bf (27)}\ f_\pi^4\ \overline{e}\Gamma_l e^c\ 
  \left(\Sigma\partial_\mu\Sigma^\dagger\right)^1_2
  \left(\Sigma\partial^\mu\Sigma^\dagger\right)^1_2
  \ +\ ...
  \nonumber\\
& = &
 -4\ C_{\beta\beta} (M_W)\ 
g^{\bf (27)}\ f_\pi^2\ \overline{e}\Gamma_l e^c\ 
 \left[ \partial_\mu\pi^-  \partial^\mu\pi^-
   \right. \nonumber\\
& & \left.
\qquad\qquad\qquad\qquad\qquad\qquad
  \ -\
   i{2\sqrt{2}\over f}
   \left(\partial_\mu\pi^-  \partial^\mu\pi^- \pi^0
     \ -\
     \partial_\mu\pi^-  \partial^\mu\pi^0 \pi^-
     \right)
  \ +\ ...
  \right]
\ \ \ ,
\end{eqnarray}
where the ellipsis denote operators involving more powers of the pion field and
also contributions from higher dimension operators.
The coefficient renormalized at the weak scale is related to the coefficient at
$\Lambda$ in the leading log approximation by
\begin{eqnarray}
   C_{\beta\beta} (M_W) & = &  C_{\beta\beta} (\Lambda)
 \left[{\alpha_s(M_W)\over \alpha_s(M_t)}\right]^{-{6\over 23}}
 \left[{\alpha_s(M_t)\over \alpha_s(\Lambda)}\right]^{-{2\over 7}}
\ \ \ ,
\end{eqnarray}
assuming that $\Lambda\gg m_t$ with $m_t$ the top quark mass.
Naive
dimensional analysis suggests that
$C_{\beta\beta} (\Lambda)\sim \Lambda^{-5}$.
In this evolution we have neglected the contributions from the electroweak
sector and only considered strong interaction corrections.

The inclusion of SU(3) breaking effects into the determination of
$g^{\bf  (27)}$ and its relation to the $\beta\beta_{0\nu}$ matrix elements
is expected to lead to modifications at the $30\% $ level.
A better estimate can be obtained from contributions
of the form $m_s\log m_s$ arising from kaon and eta meson loops in the
chiral expansion.
A straightforward calculation gives
\begin{eqnarray}
  g^{\bf (27)}_{\pi^-\pi^-} & = &
  g^{\bf (27)}_{K^+\pi^+\pi^0}
  \left[ 1 \ -\ {M_K^2\over 16\pi^2
      f_\pi^2}\log\left({M_K^2\over\Lambda_\chi^2}\right)
    \ +\ ... \ \right]
\ \ \ ,
\label{break}
\end{eqnarray}
where we have chosen to renormalize at the chiral symmetry breaking scale
$\Lambda_\chi\sim1\ {\rm GeV}$.
The ellipsis denote terms higher order in the chiral expansion, such as local
counterterms of the form $m_s$ and other such terms.
In arriving at eq.~(\ref{break}) we have used the Gell-Mann--Okubo mass
relation between the pion, kaon and eta meson masses and neglected the
pion mass.
Inserting the appropriate constants into eq.~(\ref{break}) gives an
enhancement of  $g^{\bf (27)}_{\pi^-\pi^-}$ over
$g^{\bf (27)}_{K^+\pi^+\pi^0}$
of about $12\%$.   This should be taken as an estimate only, as we expect local
counterterms to make contributions of roughly this size, as is the case for the
relation between $f_K$ and $f_\pi$.

If instead of left-handed interactions, the short-distance theory gave rise to
right-handed interactions, $\Gamma_q = \gamma_\mu (1+\gamma_5)$, then
\begin{eqnarray}
   {\cal L}_{\rm \beta\beta}^R
& \rightarrow &
 C_{\beta\beta} (M_W)\ 
g^{\bf (27)}\ f_\pi^4\ \overline{e}\Gamma_l e^c\ 
  \left(\Sigma^\dagger\partial_\mu\Sigma\right)^1_2
  \left(\Sigma^\dagger\partial^\mu\Sigma\right)^1_2
  \ +\ ...
\ \ \ ,
\end{eqnarray}
with the same value of $g^{\bf (27)}$ due to the parity invariance of the
strong interaction.

Therefore, given that we can compute $C_{\beta\beta} (\Lambda)$ in any given
theory of the short-distance physics, we can normalize the matrix element for
processes involving the pseudo-Goldstone bosons alone.
These processes are suppressed by the same strong interaction corrections
that suppress the
$\Delta I={3\over 2}$ decays of the standard model.
To be consistent with chiral symmetry, the matrix
element between pion states involves two derivatives at leading order,
and therefore for pions with vanishing four-momentum this source of
lepton number violation vanishes.
At higher orders in the chiral expansion momentum independent terms will
arise through insertions of the light quark mass matrix.

The nonleptonic decay of octet baryons proceeding via the ${\bf 27}$ representation
are described at leading order by the effective lagrange density
\begin{eqnarray}
{\cal L}_B & = &
\beta_{27} {G_F m_\pi^2 f_\pi\over\sqrt{2}} \ T^{ab}_{cd}\
\left(\xi \overline{B} \xi^\dagger\right)^c_a
\left(\xi B \xi^\dagger\right)^d_b
\ \ \ ,
\end{eqnarray}
where $\Sigma = \xi^2$.
The best fit to the available data gives $\beta_{27} \sim -0.16$,
but with large uncertainties,
as determined in \cite{Valen}.
Inserting the tensor structure for $\beta\beta_{0\nu}$ decay in
Eq.~(\ref{doubleop}),
gives a contribution to
$\Sigma^-\rightarrow \Sigma^+ e^- e^-$, but does not contribute to processes
involving the nucleons, due to isospin considerations 
(the  $\beta\beta_{0\nu}$ operator has $(\Delta I ,\Delta I_Z) = (2,+2)$).
Therefore there are no momentum independent operators involving a single
nucleon and multiple pions, in the chiral limit.
The first non-zero amplitudes involve the momentum of the pions, or insertions
of the light quark mass matrix.  The momentum dependent interactions have been
discussed previously (e.g. \cite{Wick,Vergados,Faessler,Klapdor}).
However, the present experimental determinations of the 
nonleptonic amplitudes are not precise enough to 
constrain these higher dimension operators.

In order for two s-wave nucleons to scatter via the 
$\beta\beta_{0\nu}$ interaction they must be in the $\si$-channel,
and therefore in a ${\bf 27}\oplus {\bf 8}$ of SU(3).
The effective operators that contribute to the nonleptonic and
$\beta\beta_{0\nu}$ interaction are found by forming all singlets from
$({\bf 27}\oplus {\bf 8})\otimes {\bf 27}\otimes ({\bf 27}\oplus
{\bf 8})$.
At present there are no firm determinations of any
$\Delta~I~=~{3\over 2}$, 
$\Delta s=1$
four-baryon weak interactions, and  therefore
we are unable  constrain
the coupling constants of these interactions.
More precise measurements of the properties and decay patterns of
hypernuclei would improve the situation
(e.g. \cite{hyper} and references therein).

Therefore, we have shown that the hadronic matrix elements between pion states
of operators that may contribute to
$\beta\beta_{0\nu}$ decay can be normalized to the rate for 
$K^+\rightarrow \pi^+\pi^0$
in the limit of SU(3) symmetry.
Unfortunately, the current precision of data from hyperon decays and from
hypernuclei does not allow
for constraints to be placed on the interactions  involving one
or more nucleons.

\vskip 0.1in

This work was stimulated by a nice talk given by Petr Vogel at the DNP meeting
in Santa Fe.
I would like to thank Petr Vogel and Wick Haxton for interesting discussions.
This work is supported in part by the U.S. Dept. of Energy under
Grants No. DE-FG03-97ER41014.

\end{document}